\documentclass{article}
\usepackage[utf8]{inputenc}
\usepackage{graphicx}
\usepackage[a4paper=true]{hyperref}
\usepackage{enumitem}
\usepackage[square,sort,comma,numbers]{natbib}
\usepackage{breakurl}
\title{New Nomenclature Rules\\ for Meteor Showers Adopted}
\author{Tadeusz J. Jopek$^3$, M\'{a}ria Hajdukov\'{a}$^1$, Regina Rudawska$^2$,\\ Masahiro Koseki$^4$, Gulchehra Kokhirova$^5$, Lubo\v{s} Neslu\v{s}an$^6$}
\date{\small{$^1$Astronomical Institute, Slovak Academy of Sciences, Slovakia, $^2$RHEA group/ESA ESTEC, The Netherlands, $^3$Astronomical Observatory Institute, Faculty of Physics,  A.M. University, Poznań, Poland, $^4$The Nippon Meteor Society, Japan, $^5$Institute of Astrophysics, National Academy of Sciences of Tajikistan, Republic of Tajikistan, $^6$Astronomical Institute of the Slovak Academy of Sciences, Tatranská Lomnica, Slovak Republic}}

\begin{document}
\maketitle
\noindent \small{Submitted to: New Astronomy Reviews}

\subsection*{\centering Abstract}
The Shower Database (SD) of the Meteor Data Center (MDC) had been operating on the basis of stream-naming rules which were too complex and insufficiently precise for 15 years. With a gradual increase in the number of discovered meteor showers, the procedure for submitting new showers to the database and naming them lead to situations that were inconsistent with the fundamental role of the SD - the disambiguation of stream names in the scientific literature. Our aim is to simplify the meteor shower nomenclature rules. We propose a much simpler set of meteor shower nomenclature rules, based on a two-stage approach, similar to those used in the case of asteroids. The first stage applies to a new shower just after its discovery. The second stage concerns the repeatedly observed shower, the existence of which no longer raises any doubts. Our proposed new procedure was approved by a vote of the commission F1 of the IAU in July 2022. 

\section{Introduction}

   The Meteor Data Center (MDC)\footnote{
   \url{https://www.iaumeteordatacenter.org/}.} is responsible for the unique designation of each new meteor shower. Meteor showers and their basic parameters are registered in the Shower Database (SD) of the MDC, which is used by the whole meteor science community. 
   
   The SD was created as an independent part of the IAU MDC in 2007, and was posted on the website \url{https://www.ta3.sk/IAUC22DB/MDC2007/}. Its development has been described in the following publications: \cite{2010IAUTB..27..158B, 2010IAUTB..27..177W, 2011msss.conf....7J, 2014me13.conf..353J, 2017PSS..143....3J, 2020PSS..18204821J, 2021JIMO...49..163R}  and \cite{2022AA...02..1H}. The SD of the MDC acts in conjunction with the Working Group (WG) on Meteor Shower Nomenclature of the IAU Commission F1, ”Meteors, Meteorites, and Interplanetary Dust”.
   
   Over the past fifteen years, a variety of difficulties has occurred with the upkeep of the SD due to a radical increase in reports of minor meteor showers caused by progress in video observation techniques. The main difficulty is in distinguishing between a real minor shower and a possible random grouping of sporadic meteors \citep{2019msme.book..210W}. 
   
   The definitions of a meteoroid stream and meteor shower that we currently have do not help much. In 1961, Commission 22 of the IAU (currently Commission F1) defined a meteor shower as {\it "a number of meteors with approximately parallel trajectories and a meteoroid stream as a number of meteoroids with nearly identical orbits"}, \citep[see e.g.][]{1963Metic...2....7M}. In 2018, the problem was revisited by Commission F1. New definitions were published in \cite{2016JIMO...44...31B} and are included in the IAU Transactions in electronic form.\footnote{The full text of the new \index{Meteor shower!IAU definition}definitions of notions used by meteor astronomers is available on the IAU Website:
   \url{https://www.iau.org/static/science/scientific_bodies/commissions/f1/meteordefinitions_approved.pdf}.}
   According to these definitions: {\it "a meteoroid stream is a group of meteoroids which have similar orbits and a common origin; a meteor shower is a group of meteors produced by meteoroids of the same meteoroid stream"}. 
   What one should understand by 'similar orbits' was not specified; however, with regard to different areas of orbital phase space, the specification of 'similar orbits' might be an impossible task, especially for minor meteor showers that have not yet been defined. So, the definitions are only of a general nature. 
   
   The existence of minor meteor showers has to be confirmed in a statistical way. The detection of them is, however, limited by the accuracy of the observations and/or the method used in determining the meteoroids‘ orbital data. Moreover, different methods used for the separation of minor showers from the same database may yield different results \citep[see][]{2017A&A...598A..40N}. 
   The detection limit of very minor showers may be extended by more numerous and more precise orbital data as well as by more robust statistical methods \citep{2019msme.book..210W}.
   
   It happens on a quite regular basis that a minor shower submitted to the MDC turns out to be non-existent. To deal with these difficulties and to adapt to the expected influx of reports on new meteor shower discoveries in the near future, we propose  a change in the nomenclature rules for meteor showers. \\

\section{The motivation for a new approach}
\subsection{Procedure of the naming meteor showers}
\label{procedure}
For the last 15 tears of the MDC SD, each new meteor shower has been given a unique name, and both a numerical and a 3-letter code, at the moment of providing the shower data to the MDC. 
The discoverer of meteor shower proposed a new, unique name, according to the shower nomenclature rules. The suggested name was verified by the MDC.
If needed, the MDC corrected the name and assigned the two codes to the new shower.
Unfortunately as a result of the application of these meteor shower nomenclature rules, the obtained shower names were lengthy, and in order to establish them and check their uniqueness, quite complex software should be used.

After this process, the data of the new stream were placed on the Working List of the MDC SD. Subsequently, the discoverer of the stream had six months to provide a copy of the publication to the MDC, describing their discovery. 
The MDC SD team accepts meteor showers published in all peer-reviewed scientific journals in the field of astronomy, as well as in the two journals dedicated to amateur meteor astronomy: WGN, the Journal of the IMO, and Meteor News. 

If the publication does not reach the MDC on time, the stream codes, name and parameters are deleted from the database. The name and codes released in this way - as they have not been published anywhere - can be reused \citep[see][]{2020PSS..18204821J}. 
However, if the publication reaches the MDC on time, a new shower with assigned codes and name, may be on the Working List forever; at least in theory. Its codes and name cannot be assigned to another shower.

Thus, the procedure for assigning names and codes to new showers was a one-step process. This was because the following step was the official recognition by the IAU only of the previously fixed name, and applied to those streams which were sufficiently well-established.

The described one-step procedure for determining the final names of showers, in the light of our experience in handling the MDC, turned out to be inadequate. This procedure made it difficult or even impossible for the MDC to fulfil its primary role - ensuring the uniqueness of the names of the meteoroid streams in the scientific literature, i.e. the names that meet the applicable meteoroid stream naming rules.
 
\subsection{Nomenclature rules --- state until July 2022}
In 2006, for the first time, the traditional rules for the naming of meteoroid streams were formalised, \citep[see][]{2007IAUBull..J,2009JIMO...37...19J, 2011msss.conf....7J}. 
Over time, based on the experience gained in the administration of the MDC SD, they were modified in accordance with the recommendations of the Working Group on Meteor Shower Nomenclature.
It was decided that the new nomenclature rules will apply to new streams. For major showers which have been known for many years, their traditional names will be used, even if they do not follow the naming rules. However, the names of those showers which were named after the parent body, e.g. Giacobinids, Halleids, are no longer acceptable.
\\

New meteor showers (meteoroid streams) may be named according to the following general rule:
\[
 [N,S]\,[D]\,[M]\,[S]-[Constellation] \, [C]   
\]

The {\it Constellation} component is mandatory, the others are optional. The order of the components is also mandatory. 

The naming of a meteor shower starts with the mandatory component [constellation]. If the created name is not unique in the MDC SD, the non-mandatory components (one by one) are added in the order given until the name thus created is unique. The use of the individual components is as follows:   
\begin{enumerate}
\itemsep 2pt
\item {[{\it Constellation}]}. A new meteor shower (and a meteoroid stream) should be named after the constellation that contains the star nearest to the shower radiant. The name should be composed of the possessive Latin name of the constellation. If the Latin name contains one of the following suffixes: {\it ae, is, i, us, ei, ium,} or {\it orum}  --- it should be replaced by {\it id}, or plural {\it ids}. For example, meteors with radiants in Aquarius are Aquariids, in Orion are Orionids, in Ursa Major are Ursae Majorids.
However, in the case of the constellation of Hydrus, the meteors are called Hydrusids, in order to avoid confusion with meteors radiating from the constellation of Hydra.
\begin{enumerate} 
\item When the name of a constellation consists of two parts, e.g. Canes Venatici (Latin possessive Canum Venaticorum), then the shower name is Canum Venaticids. Only the second part of the constellation's possessive  form is modified by {\it ids}.
\item When the activity of a shower extends over the boundary of two constellations, its name may contain the corresponding Latin forms of both constellations connected by a dash, e.g. Cepheids-Cassiopeiids, Puppids-Velids, Canum Venaticids-Bootids.\itemsep 0pt
\end{enumerate}
\itemsep 0pt
\end{enumerate}
The remaining options are as follows:
\begin{enumerate}[resume]
\itemsep 2pt
\item {[{\it S}]}. If there is more than one radiant in a constellation then, additionally, the designation of the star (designated in Bayer or Flamsteed systems) nearest to the radiant is assigned to the shower name. A lowercase Greek or Latin letter is used in the Bayer designation system, a number in the Flamsteed designation system. To avoid confusion, the two components should be separated by a dash, e.g. alpha-Capricornids, 49-Andromedids.  
\begin{enumerate}\itemsep 0pt 
\item If a meteor shower radiant is near the border of a constellation and the nearest star is in the neighbouring constellation, then the shower is named after that star and constellation.
\end{enumerate}

\item {[{\it M}]}. If necessary, one may add the name of the month of the shower activity so that showers from the same constellation can be distinguished. The name of the month may appear immediately before the Latin form of the constellation name, e.g. October Draconids, or before the Bayer or Flamsteed star designation, e.g. July mu-Serpentids, October 6-Draco\-nids.
\begin{enumerate}
    \item  When the shower activity extends over two consecutive months, the name of the stream may contain the names of both months separated by a dash, e.g. September-October Lyncids.
\end{enumerate}
\item {[{\it D}]}. For daytime showers, it is customary to add 'Daytime' to the Latin form of the constellation name, e.g. Daytime Arietids, or in more complex cases: Daytime kappa-Librids, Daytime April Cetids or Daytime June alpha-Orionids. To be called a daytime shower, its radiant should be at an angular distance less than 32 degrees from the Sun center when the meteors are observed.
\item {[{\it N, S}]}. If necessary, the names of the shower may be supplemented, on the left, by the words 'Northern' and 'Southern'. This refers to cases where meteoroid streams (meteor showers) originated from the same parent body and have radiants situated north or south of the ecliptic plane. Acceptable names of a shower are, for example, Northern Taurids, Southern Daytime May Arietids, Northern December omega-Ursae Majorids, etc.
\item {[{\it C}]}. In the case of groups of showers, e.g. originated from the same parent body, the name of the group is created analogically to a single shower, but supplemented from the right with the word Complex, e.g. delta Aquariid Complex, Daytime May Arietid Complex, Southern Librids-Lupids Complex, etc. In order to better distinguish more complex groups, names containing the Roman numerals 'I', 'II'' ... preceding the word Complex are also allowed, e.g. Centaurid II Complex, Puppid-Velid I Complex, etc.
\itemsep 0pt
\end{enumerate}

All names of new showers proposed by the discoverer are accepted in the IAU Meteor Data Center Shower Database if they are unique, i.e. if they do not exist in the MDC, and if they conform to naming rules. As additional examples, we give here the simplest and the most extensive names of showers from the MDC: Orionids and Northern December omega-Ursae Majorids. 

%
\subsection{Weaknesses of the old naming rules}
Although all meteor shower naming rules, as well as their modifications, were discussed by WG members and approved by the Commission 22, the predecessor of the IAU Commission F1, they were not posted on the official IAU website for "Naming of Astronomical Objects" until this year. In our opinion, such a situation was highly uncomfortable for meteor astronomers. 

Often, meteor astronomers themselves have not respected the naming rules they set. 
In addition, some of these rules were not unambiguous; hence, they were only general in nature. During the Meteoroids 2013 conference in Poznań, \citet{2014me13.conf..353J} showed that, among the $646$ shower radiants available at the time, more than $50$\% had names incompatible with the nomenclature rules. The inconsistencies mainly concerned the name of the constellation and the name of the star closest to the shower radiant. 

The ambiguity relating to the rules listed above arises from various issues.
For example, to verify whether the second rule has been correctly applied, it would be necessary to establish a star catalogue and the limiting star magnitude in advance. Both issues are not formally established in the naming rules set.

\begin{figure}[ht]
\centerline{
\hbox{
\includegraphics[width=0.9\textwidth]{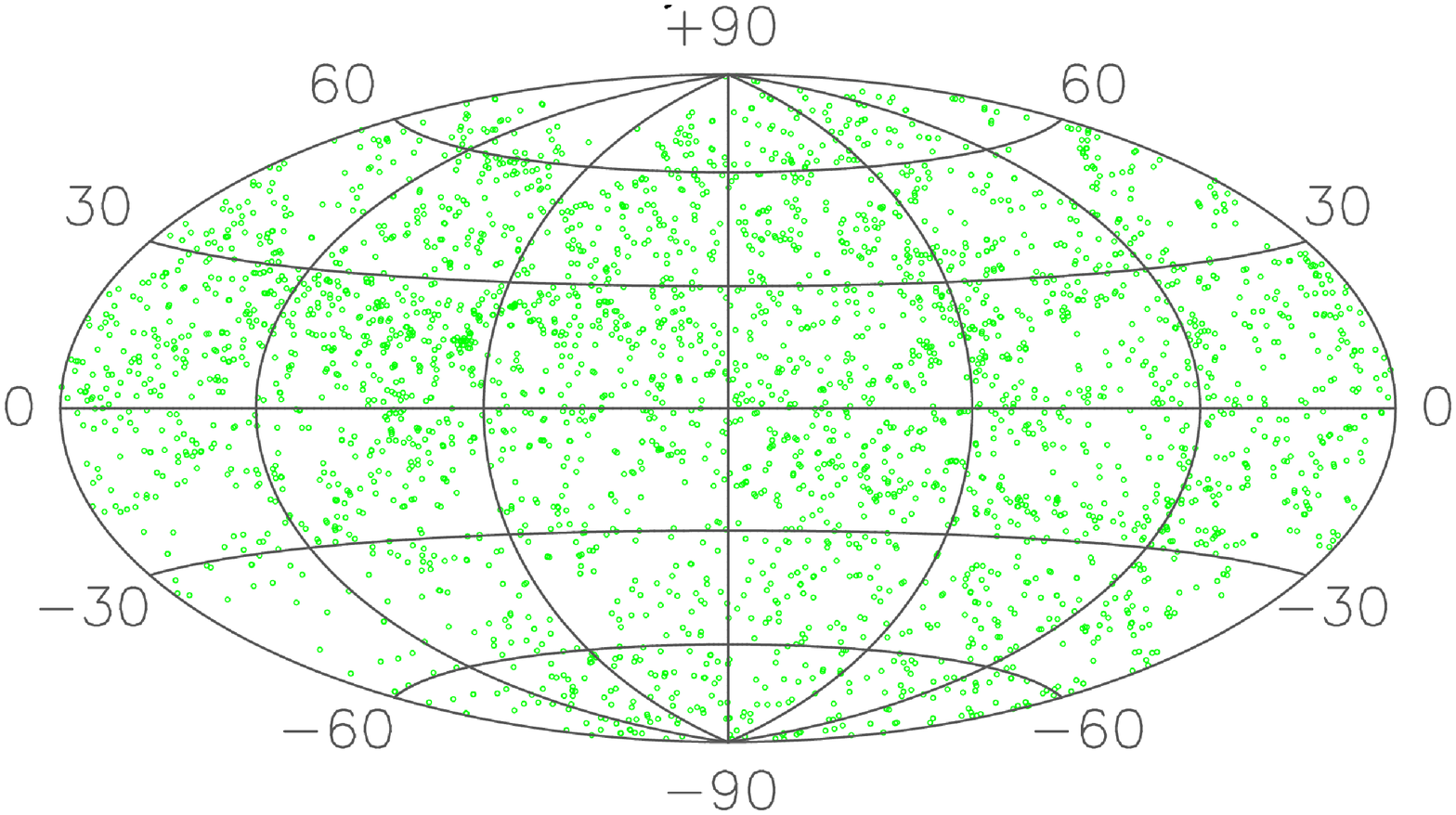}
}
}
\caption{Hammer-Aitoff diagram of the equatorial coordinates of $3141$ Bayer and/or Flamsteed stars taken from the Bright Star Catalogue. The~sky coverage on this diagram is not perfect; one can clearly see a few regions significantly less populated by stars. In such regions it is much more difficult to establish a meaningful stream name than in regions with a large number of stars.}
\label{fig:bsc}
\end{figure}

For our convenience, in the MDC SD, we chose a subset of the~Yale Bright Star Catalogue (BSC), 5th Revised Ed. \citep{1991bsc..book.....H}.\footnote{
The BSC is available free of charge at \url{http://tdc-www.harvard.edu/catalogs/bsc5.html}.}
$3141$ stars were selected from the~BSC, all stars brighter than $6.5$ magnitude for which Bayer and/or Flamsteed names were available \citep[see][]{2014me13.conf..353J}. The distribution of these stars on the celestial sphere is shown in Figure \ref{fig:bsc}.
However, as practice has shown, our choice did not suit all users, especially those users who named streams based on a different star catalogue. It is clear that, without adopting a common star catalogue, it will not be easy, and sometimes even impossible, to establish the correctness of the proposed stream name.

Other problems arise due to imprecise terms used in the rules, i.e. 'near the border'. There is no agreement as to exactly what one should understand by 'near'. 

Another reason for changing the nomenclature rules is the problem of ensuring the uniqueness of the shower names in the scientific literature. During many years with the SD, we have had approximately 400 cases of streams that were named and spent several months on the MDC web Working List, but which did not meet the publication conditions or their submissions were withdrawn by the providers. Their names and codes could have been used again for a different stream and would have thus created confusion. 
Some showers have been proved to be duplicates, and a number of such cases is expected to grow. The duplicate parameters have been added as another solution to another already existing shower. However, the earlier name of the duplicate shower still functions in the literature. 

In $208$ cases, the meteor showers data have been moved to the List of Removed Showers for various reasons, see \cite{2022AA...02..1H}. Their names and codes cannot be reused.

Thus, we can clearly see that determining the final names of showers at the time of their submission has been causing confusions in the literature and/or blocking the possibility of using many names in more justified cases. 

Making the old shower-naming rules more precise would have been, of course, possible, but then they would have been less and less transparent and more difficult to apply and be verified. The procedure for the naming of meteor showers (meteoroid streams) should be a two-step process, as is the case with the naming of comets or asteroids. 
The first would apply to new showers which do not yet have names and codes; the second to established showers approved by the IAU.
%
\section{New nomenclature rules for meteor showers}

 We proposed the following two-stage procedure for naming meteor showers: 
\begin{enumerate}
\item[] \textbf{Stage one} -- new meteor showers sent to the MDC are given a provisional designation only (not a name).
The provisional designations are based on the date of submission and are assigned by the MDC according to a rule that involves:
\begin{enumerate}[label={\arabic*.}]
\item the prefix M,
\item the year of discovery followed by a hyphen, and an uppercase letter identifying the half-month of observation during that year; A for the first half of January, B for the second half, and so on; see Table \ref{tab:halfmonth},
\item a number representing the order of the shower submission to the MDC within that half month.
\end{enumerate} 

Thus, for example, the third shower recently submitted to the MDC, on May 29 2022, would be designated as M2022-K3.\\

\item[] \textbf{Stage two} -- when a meteor shower has become well confirmed (its regular activity, origin, etc.) and meets the required criteria for 'established status' (see \cite{2022AA...02..1H}, Sect.~3) 
it will be given a final designation according to the following schema:  
\begin{enumerate}[label=\arabic*.]
\item a prefix M followed by a hyphen,
\item the IAU MDC numerical code (a number issued sequentially by the MDC),
\item a name (The discoverer is invited to propose a name for their stream. All proposed names are judged by the Working Group on Meteor Shower Nomenclature of the IAU.)
\end{enumerate}
 The final shower designation and the name will be officially approved by the IAU.
\end{enumerate}
Thus, in the above-mentioned example: the final designation would be M-1192 and e.g. the name alpha-Virginids.\\

The new procedure for naming meteoroid streams will be consistent with those already used for the naming of comets and asteroids. 
Moreover, due to the possibility of observing meteor showers outside the Earth’s atmosphere, e.g. in the atmosphere of Mars, in all similar cases, a new naming procedure would be very desirable.
\begin{table}[!ht]
\caption{The relationship between the letters of the Latin alphabet and the half-month intervals in the calendar year used for the designation of the meteor showers. Notice. Letters I and Z are unused.}
\small
\begin{center}
\begin{tabular}{cccc}
\hline
\centering
{Latin letter} &   {Half month} & {Latin letter} & Half month  \\
\hline
    A    &  Jan. 1-15    &       B   &    Jan. 16-31 \\
    C    &  Feb. 1-15    &       D   &    Feb. 16-29\\
    E    &  Mar. 1-15    &       F   &    Mar. 16-31\\
    G    &  Apr. 1-15    &       H   &    Apr. 16-30\\
    J    &  May  1-15    &       K   &    May  16-31\\
    L    &  June 1-15    &       M   &    June 16-30\\
    N    &  July 1-15    &       O   &    July 16-31\\
    P    &  Aug. 1-15    &       Q   &    Aug. 16-31\\
    R    &  Sept.1-15    &       S   &    Sept.16-30\\
    T    &  Oct. 1-15    &       U   &    Oct. 16-31\\
    V    &  Nov. 1-15    &       W   &    Nov. 16-30\\
    X    &  Dec. 1-15    &       Y   &    Dec. 16-31\\
\hline
\end{tabular}
\end{center}
\normalsize
\label{tab:halfmonth}
\end{table}
%
%

We introduced this new approach for the meteor shower nomenclature at the Meteoroids conference 2022, discussed it with members of the WG and with the participants of the 2022 June 15 Business meeting of the F1 Commission. It was approved by a majority of members of the IAU Commission F1 who participated in the electronic voting completed on July 20, 2022.

The new naming rules of the meteor showers are given on the website  \url{https://www.ta3.sk/IAUC22DB/MDC2022/Dokumenty/shower_nomenclature.php}.
They also were posted on the official IAU website 'Naming of Astronomical Objects' \url{https://www.iau.org/public/themes/naming/}. 

\section*{Conclusions}
 We proposed a two-step procedure for the designation and naming of meteor showers. The first stage will apply to a new shower just after its discovery. Each new meteoroid stream sent to the MDC will be given a provisional designation only (not a name). The second stage will apply to a repeatedly observed shower, the existence of which no longer raises any doubts. Each shower that meets the required criteria for 'established status' will be given a final designation by the MDC and a name suggested by the discoverer. The cross-reference between the final and provisional designations will be provided at the MDC.

\section*{Acknowledgements}
The authors like to acknowledge Members of the Commission F1 IAU for their discussion and remarks related with either the new meteor nomenclature rules as well as the criteria for nominating meteor showers for established status.
The work was, in part, supported by the VEGA – the Slovak Grant Agency for Science, grant No. 2/0009/22.

This research has made use of NASA's Astrophysics Data System Bibliographic Services.

\bibliographystyle{unsrt}
\bibliography{IAUMDC_naming_rules_Arxiv} 

\begin{thebibliography}{10}

\bibitem{2010IAUTB..27..158B}
Edward L.~G. {Bowell}, Karen~J. {Meech}, Iwan~P. {Williams}, Alan {Boss},
  R{\'e}gis {Courtin}, Bo~{\r{A}}.~S. {Gustafson}, Anny-Chantal
  {Levasseur-Regourd}, Michel {Mayor}, Pavel {Spurn{\'y}}, Junichi {Watanabe},
  Guy~J. {Consolmagno}, Julio~A. {Fern{\'a}ndez}, Walter~F. {Huebner},
  Mikhail~Ya. {Marov}, Rita~M. {Schulz}, Giovanni~B. {Valsecchi}, and Adolf~N.
  {Witt}.
\newblock {Division III: Planetary Systems Science}.
\newblock {\em Transactions of the International Astronomical Union, Series B},
  6(T27):158--167, May 2010.

\bibitem{2010IAUTB..27..177W}
Junichi {Watanabe}, Peter {Jenniskens}, Pavel {Spurn{\'y}}, Ji{\v{r}}{\'\i}
  {Borovi{\v{c}}ka}, Margaret {Campbell-Brown}, Guy {Consolmagno}, Tadeusz
  {Jopek}, Jeremie {Vaubaillon}, Iwan~P. {Williams}, and Jin {Zhu}.
\newblock {Commission 22: Meters, Meteorites and Interplanetary Dust}.
\newblock {\em Transactions of the International Astronomical Union, Series B},
  6(T27):177--179, May 2010.

\bibitem{2011msss.conf....7J}
T.~J. {Jopek} and P.~M. {Jenniskens}.
\newblock {The Working Group on Meteor Showers Nomenclature: A History, Current
  Status and a Call for Contributions}.
\newblock In W.~J. {Cooke}, D.~E. {Moser}, B.~F. {Hardin}, and D.~{Janches},
  editors, {\em Meteoroids: The Smallest Solar System Bodies}, pages 7--13,
  July 2011.

\bibitem{2014me13.conf..353J}
T.~J. {Jopek} and Z.~{Ka{\v{n}}uchov{\'a}}.
\newblock {Current status of the IAU MDC Meteor Showers Database}.
\newblock In T.~J. {Jopek}, F.~J.~M. {Rietmeijer}, J.~{Watanabe}, and I.~P.
  {Williams}, editors, {\em Meteoroids 2013}, pages 353--364, July 2014.

\bibitem{2017PSS..143....3J}
T.~J. {Jopek} and Z.~{Ka{\v{n}}uchov{\'a}}.
\newblock {IAU Meteor Data Center-the shower database: A status report}.
\newblock {\em Planetary Space Science}, 143:3--6, September 2017.

\bibitem{2020PSS..18204821J}
Peter {Jenniskens}, Tadeusz~J. {Jopek}, Diego {Janches}, Maria {Hajdukov{\'a}},
  Gulchehra~I. {Kokhirova}, and Regina {Rudawska}.
\newblock {On removing showers from the IAU Working List of Meteor Showers}.
\newblock {\em Planetary Space Science}, 182:104821, March 2020.

\bibitem{2021JIMO...49..163R}
Regina {Rudawska}, Maria {Hajdukova}, Tadeusz~J. {Jopek}, Lubos {Neslusan},
  Marian {Jakubik}, and Jan {Svoren}.
\newblock {Status of the IAU Meteor Data Center}.
\newblock {\em WGN, Journal of the International Meteor Organization},
  49(6):163--168, December 2021.

\bibitem{2022AA...02..1H}
M.~{Hajdukov{\'a}}, R.~{Rudawska}, T.J. {Jopek}, M.~{Koseki}, G.~{Kokhirova},
  and L.~{Neslu{\v{s}}an}.
\newblock {Modification of the Shower Database of the IAU Meteor Data Center}.
\newblock {\em submitted to AA}, October 2022.

\bibitem{2019msme.book..210W}
Iwan~P. {Williams}, Tadeusz~J. {Jopek}, Regina {Rudawska}, Juraj {T{\'o}th},
  and Leonard {Korno{\v{s}}}.
\newblock {\em {Minor Meteor Showers and the Sporadic Background}}, page 210.
\newblock Cambridge University Press, 2019.

\bibitem{1963Metic...2....7M}
P.~M. {Millman}.
\newblock {Terminology in Meteoritic Astronomy}.
\newblock {\em Meteoritics}, 2:7--10, 1963.

\bibitem{2016JIMO...44...31B}
J.~{Borovi{\v{c}}ka}.
\newblock {About the definition of meteoroid, asteroid, and related terms}.
\newblock {\em WGN, Journal of the International Meteor Organization},
  44(2):31--34, April 2016.

\bibitem{2017A&A...598A..40N}
L.~{Neslu{\v{s}}an} and M.~{Hajdukov{\'a}}.
\newblock {Separation and confirmation of showers}.
\newblock {\em Astronomy \& Astrophysics}, 598:A40, February 2017.

\bibitem{2007IAUBull..J}
Peter {Jenniskens}.
\newblock {The IAU Meteor showers Nomenclature rules}.
\newblock In {\em IAU Information Bulletin 99}, pages 60--62, January 2007.

\bibitem{2009JIMO...37...19J}
P.~{Jenniskens}, T.~J. {Jopek}, J.~{Rendtel}, V.~{Porub{\v{c}}an},
  P.~{Spurn{\'y}}, J.~{Baggaley}, S.~{Abe}, and R.~{Hawkes}.
\newblock {On how to report new meteor showers}.
\newblock {\em WGN, Journal of the International Meteor Organization},
  37(1):19--20, February 2009.

\bibitem{1991bsc..book.....H}
Dorrit {Hoffleit} and Carlos {Jaschek}.
\newblock {\em {The Bright star catalogue}}.
\newblock Yale University, 1991.

\end{thebibliography}
\end{document}